\documentclass{article}
\usepackage{spconf,amsmath,graphicx}
\usepackage{url}
\usepackage{amssymb}
\usepackage{multirow}
\usepackage{caption}
\usepackage{subcaption}
\usepackage{color}
\usepackage{tikz}
\usepackage{booktabs}

\title{Learning Audio Embeddings with User Listening Data \\ for Content-based Music Recommendation}
\name{Ke Chen $^1$, Beici Liang $^2$, Xiaoshuan Ma $^2$, Minwei Gu $^2$}
\address{
         $^1$   CREL, Music Department, University of California San Diego, USA\\
	            knutchen@ucsd.edu\\
	     $^2$   QQ Music BU, Tencent Music Entertainment, China\\ 
                \{beiciliang, stanfordma, torogu\}@tencent.com
	     }

\begin{document}
%
\maketitle

\begin{abstract}

Personalized recommendation on new track releases has always been a challenging problem in the music industry. To combat this problem, we first explore user listening history and demographics to construct a user embedding representing the user's music preference. With the user embedding and audio data from user's liked and disliked tracks, an audio embedding can be obtained for each track using metric learning with Siamese networks. For a new track, we can decide the best group of users to recommend by computing the similarity between the track's audio embedding and different user embeddings, respectively. The proposed system yields state-of-the-art performance on content-based music recommendation tested with millions of users and tracks. Also, we extract audio embeddings as features for music genre classification tasks. The results show the generalization ability of our audio embeddings.

\end{abstract}
\begin{keywords}
	Audio Embedding, Music Representations Learning, Music Recommendation.
\end{keywords}

\section{Introduction}
With the increasing number of tracks in online streaming services, a personalized music recommendation system plays a vital role in discovering potential tracks and distributing them to the right users. Collaborative Filtering (CF) is a commonly used method that can infer similarities between items for recommendation \cite{sarwar2001item}. It can be formulated as a deep neural network in \cite{wang2015collaborative} to model user–item interactions and offer better recommendation performance. Using such a recommendation system as in \textit{YoutubeDNN} \cite{youtubednn}, a User Embedding (UE) can be learned as a function of the user’s history and context.

While CF performs well when historical data are available for each item, it suffers from the cold start problem for novel or unpopular items. New track releases will not be recommended unless their similarities can be learned directly from the audio content. This has motivated researchers to improve content-based recommendation system, which operates on Audio Embedding (AE). Here AE refers to music representations that are extracted from audio. For example, the AE in \cite{bogdanov2013semantic} corresponds to probabilities of 62 semantic descriptors which include musical genre, mood, instruments and so on. These probabilistic estimates are the output of Support Vector Machines (SVMs) \cite{svm}, which operate on low-level audio features. In \cite{van2013deep, cf-m-3}, the use of deep Convolutional Neural Networks (CNN) to predict AE is investigated. It outperforms traditional approaches using a Bag-of-Words representation from feature engineering. Recently, metric learning \cite{xing2003distance} has been used to learn AE. In \cite{ct-3}, same/different-artist track pairs are used as supervision. And in \cite{cf-m-4}, metadata and co-listen statistics are firstly used to train AE from audio inputs for music recommendation and tagging. Based on this work, attribute embeddings are obtained for playlist exploration and expansion in \cite{Patwari2020semantically}.

However, user listening data have not been fully utilized in content-based music recommendation. Since most of the existing studies \cite{van2013deep, ct-2} rely on the Echo Nest Taste Profile \cite{mcfee2012million} and the Million Song Dataset \cite{Bertin-Mahieux2011}, user listening data only include play counts associated with limited users and tracks. Still, such data have shown how they surpass using audio only in mood classification \cite{korzeniowskimood} and estimation of contextual tags \cite{ibrahim2020should}. We believe a more informative UE can be obtained if more user listening data can be included. Such UE can also work together with AE to extract information that may not necessarily present in the audio content.

In this paper, with real-world data collected from an online music platform, we propose a model with two branches as presented in Section \ref{sec:model}. One is the \textit{user branch} that considers different aspects of user listing data to train UE. This UE is further utilized in the \textit{audio branch} that uses metric learning with user-liked/disliked track pairs to obtain AE of each track. With the trained model, accurate and efficient representations of tracks can be obtained for music recommendation and genre classification. Experimental results in Section \ref{sec:exp} demonstrate significant improvements in related tasks.

\section{Proposed Model}
\label{sec:model}
We first detail the user branch to obtain UE, and then present how the UE is used in the audio branch in a metric learning framework to obtain AE for content-based music recommendation. 

\subsection{User Branch}
\begin{figure}
    \centering
    \includegraphics[width=\columnwidth]{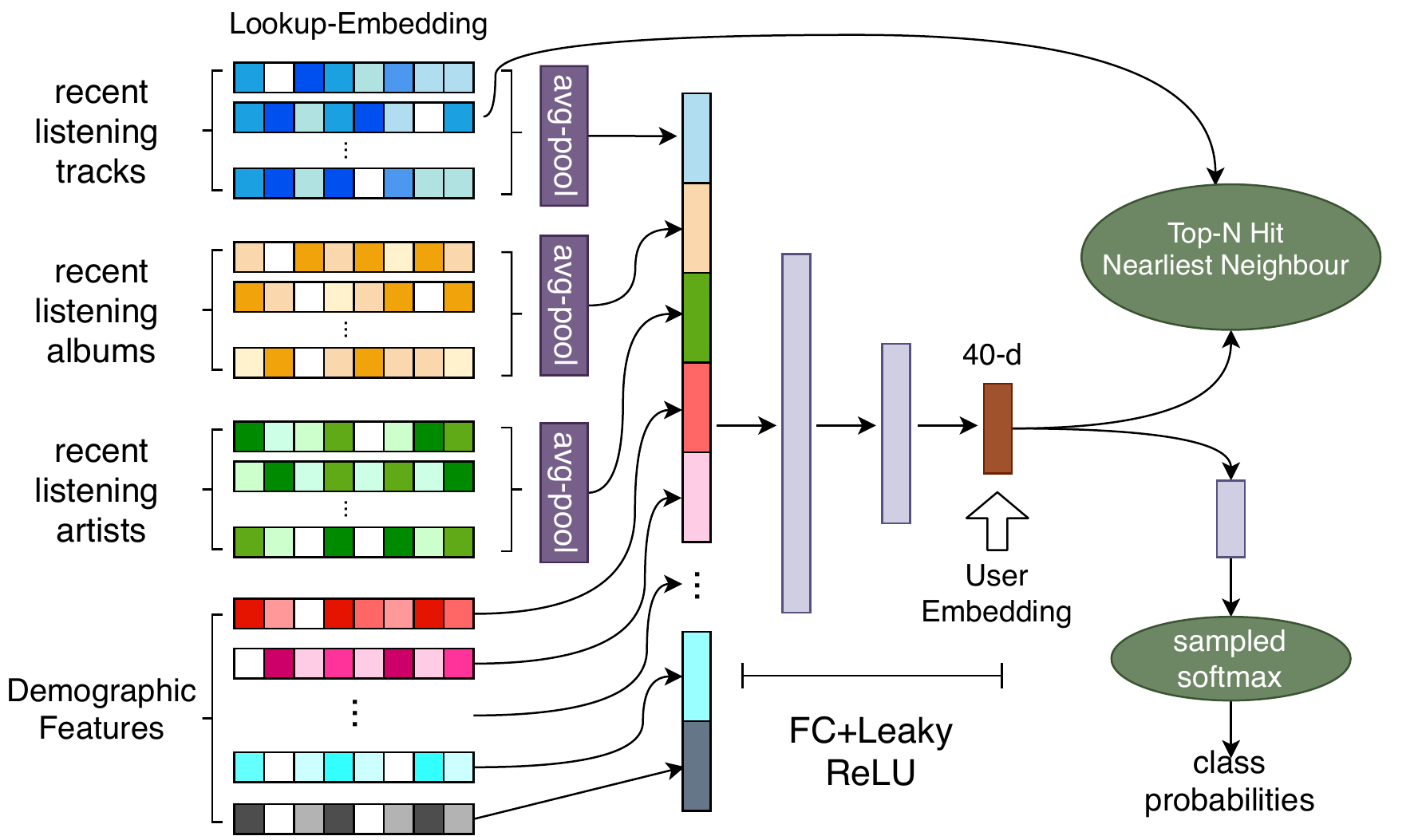}
    \caption{Architecture of the user branch to obtain UE using users' listening history and demographics.}
    \label{fig:user-branch}
\end{figure}
The user branch of our model is to encode user's music preference according to the user's listening history and demographics. It is designed to address a classification problem formulated as follows: given a user's listening history $X_{1:t-1}=\{x_1, x_2, ..., x_{t-1}\}$ and demographics $D$, the model can classify $x_t$ as a user-liked/disliked track by maximizing the conditional probability:
\begin{equation}
\begin{split}
&\max P(x_t | X_{1:t-1}, D) \\
&=P(x_t | x_1,..x_{t-1},a_1,...,a_{t-1},s_1,..,s_{t-1},D)
\end{split}
\end{equation}
where $x_t$ denotes music tracks, $a_t$ denotes albums in which these tracks appear, and $s_t$ denotes the corresponding artists. All the data are represented by IDs which can be mapped to lookup-embeddings.

As shown in Figure \ref{fig:user-branch}, the network of the user branch is inspired by the structure of \textit{YoutubeDNN} \cite{youtubednn}. To accommodate music recommendation tasks, we use both explicit and implicit feedback of tracks to train the network. For instance, a user putting a track in a playlist is a positive example. To feed all lookup-embeddings to the network, embeddings of several groups are averaged into fixed-width vectors of 40 dimensions. They are concatenated into a vector, which is passed into hidden layers that are fully connected with leaky-ReLU activation. In training, a cross-entropy loss is minimized for the positive and the sampled negative classes.

The final state before the output layer is used as UE with a dimension of 40. At serving, nearest-neighbor lookup can be performed on UE to generate top-N candidates for music recommendation.

\subsection{Audio Branch}
\begin{figure}
    \centering
    \includegraphics[width=\columnwidth]{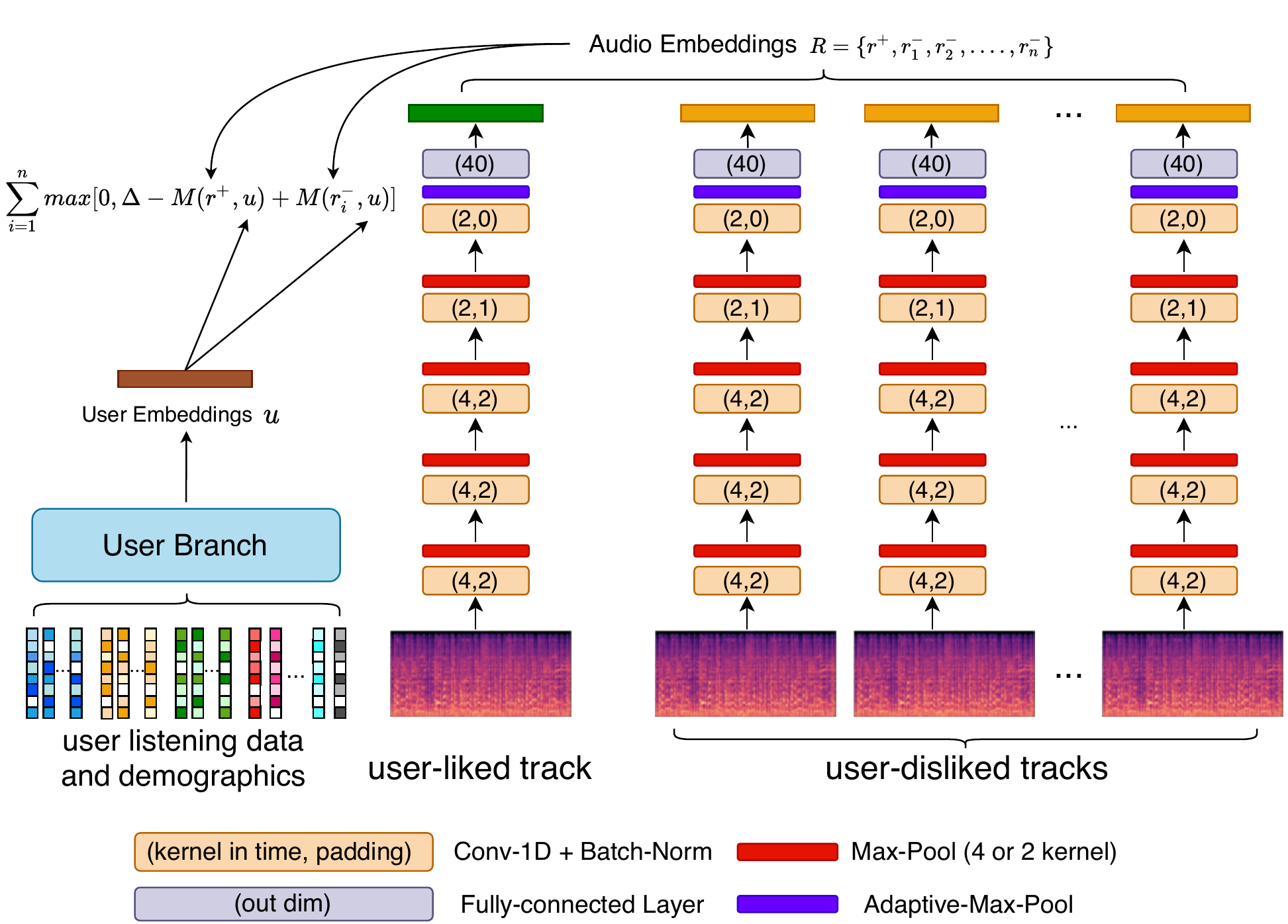}
    \caption{Architecture of the audio branch to obtain AE using UE and audio data.}
    \label{fig:audio-branch}
\end{figure}
The audio branch contains a Siamese model following the architecture in \cite{ct-2} to learn AE directly from audio. It also uses UE from the user branch as an important anchor for metric learning as shown in Figure \ref{fig:audio-branch}: given a user's UE ($u$), an AE of a track liked by this user ($r^+$), and $n$ AEs of tracks disliked by this user ($\{r^-_1, r^-_2,...r^-_n\}$), the model should differ $r^+$ from $r^-$.

To get AE, we feed log-mel spectrograms of the audio inputs to the Siamese model, consisting of CNNs that share weights and configurations. As depicted in Figure \ref{fig:audio-branch}, each CNN is composed of 5 convolution and max-pooling layers. ReLU activation layer is used after every layer except for the final feature vector layer to get AE with a dimension of 40. 

In training, we apply the negative sampling technique over the pairwise similarity score, which is measured as:
\begin{equation}
    M(r,u) = \frac{u^{T}\cdot r}{\left | u \right | \cdot  \left | r \right |}
\end{equation}
The loss is then calculated using the above relevance scores obtained from the UE and the AEs of liked and disliked tracks. We use the max-margin hinge loss to set margins between positive and negative examples. This loss function is defined as follows:
\begin{equation}
    L(u,R)=\sum_{i=1}^{n}max[0, \Delta - M(r^+, u) + M(r^-_i, u)]
\end{equation}
where $\Delta$ is a margin hyper-parameter, which is set to 0.2 in this paper. 

Using the trained CNN in the audio branch, a novel track can be represented by its AE. If the similarity between the AE and a user's UE yields a high score, the track can be a candidate that fits this user's music preference. Therefore AE can be used to recommend novel or unpopular tracks for which collaborative filter data is unavailable. Experimental results verify that the proposed audio branch outperforms competing methods for content-based music recommendation.



\section{Experiments}
\label{sec:exp}

\subsection{Experiment Setup and Dataset}
Our deep collaborative filtering model in the user branch has been used to generate candidates for recommendation at QQ Music\footnote{\url{https://y.qq.com}}. It was trained with over 160 millions of user-track interaction data associated with 2 million tracks. To measure the performance, hit rate was used. The model can reach 31\% hit rate among the top-200 recommendations. A complete experimental report of the user branch is beyond the scope of this paper. We collected UE from the trained user branch for the following experiments on the audio branch.

For the metric learning in the audio branch, we created a dataset using user-liked/disliked tracks. Each user has at least 10 liked and 10 disliked tracks. Every track was processed into a log-mel spectrogram with 128 bins using a sample rate of 22050 Hz and a hop size of 512. For training efficiency, we randomly selected a segment of the log-mel spectrogram within the first 30 seconds, and denoted this segment as context duration. Intuitively, we believe a user can decide to skip or keep listening to a track in 30 seconds. In total, our dataset includes approximately 1.9 million users and 1.1 million tracks. 

Following the training configurations in \cite{ct-2}, the loss was optimized using stochastic gradient descent with 0.9 Nesterov momentum with 1e-6 learning rate decay. The initial learning rate was set to 0.1. We used Pytorch 1.5.1\footnote{\url{https://pytorch.org}} on NVIDIA V100 GPU machines for training all the models in the audio branch.

To evaluate our proposed model, we first conducted experiments on music recommendation tasks to predict a track will be liked/disliked by a given user in Section \ref{sec:eval_rec}. With a 60\%/20\%/20\% split for training/validation/testing, AUC (Area Under Receiver Operating Characteristic) and precision are used as evaluation metrics. Using the best-performed model to obtain AE, we applied the AE as features to genre classification in Section \ref{sec:eval_genre}. This helps us explore if our AE is capable of extracting certain genre information about a given track.

\begin{table}[t]
    \centering
    \begin{subtable}[t]{0.8\linewidth}
    \centering
    \begin{tabular}{@{}|l|c|c|@{}}
    \toprule
    \textbf{Model} & \textbf{Precision} & \textbf{AUC} \\ 
    \midrule
    \begin{tabular}[c]{@{}l@{}} \texttt{Basic-Binary} \end{tabular}        & 0.677         & 0.747        \\ \midrule
    \begin{tabular}[c]{@{}l@{}} \texttt{DCUE-1vs1} \\ \end{tabular} & 0.623 & 0.675 \\ \midrule
    \begin{tabular}[c]{@{}l@{}} \texttt{Multi-1vs1} \end{tabular}  & 0.745 & 0.752 \\ \midrule
    \begin{tabular}[c]{@{}l@{}} \texttt{Multi-1vs4} \end{tabular} & 0.687 & 0.749 \\ \midrule
    \begin{tabular}[c]{@{}l@{}} \texttt{Metric-1vs1} \end{tabular} & 0.691 & 0.765 \\ \midrule
    \begin{tabular}[c]{@{}l@{}} \texttt{Metric-1vs4} \end{tabular} & 0.681 & 0.778\\ \bottomrule
    \end{tabular}
    \caption{context duration equals 3 seconds.}
    \end{subtable}
     
    \begin{subtable}[t]{0.8\linewidth}
    \centering
    \begin{tabular}{@{}|l|c|c|@{}}
    \toprule
    \textbf{Model}  & \textbf{Precision} & \textbf{AUC} \\ \midrule
    \begin{tabular}[c]{@{}l@{}} \texttt{Basic-Binary} \end{tabular}        & 0.696         & 0.762        \\ \midrule
    \begin{tabular}[c]{@{}l@{}} \texttt{DCUE-1vs1} \end{tabular} & 0.644 & 0.697 \\ \midrule
    \begin{tabular}[c]{@{}l@{}} \texttt{Metric-1vs1} \end{tabular}  & 0.717 & 0.788 \\ \midrule
    \begin{tabular}[c]{@{}l@{}} \texttt{Metric-1vs4} \end{tabular} & 0.701 & 0.792 \\ \bottomrule
    \end{tabular}
    \caption{context duration equals 10 seconds.}
    \end{subtable}
    \caption{Evaluation results on the sub-dataset with different context duration.}
    \label{subset-result}
\end{table}

\subsection{Evaluation on Music Recommendation Tasks}
\label{sec:eval_rec}

\subsubsection{Models}
All the models we evaluated are introduced as follows:
\begin{itemize}
  \item \texttt{Basic-Binary}: The audio branch only contains one group of CNNs that uses either a user-liked or user-disliked track as input. Then the similarity score is passed to a sigmoid function for binary classification.
  \item \texttt{DCUE-1vs1}: This refers to the Deep Content-User Embedding Model in \cite{ct-2}, which uses user-index to obtain lookup-embeddings instead of UE in the user branch. As for the audio branch, an user-liked/disliked track pair is used as audio inputs, i.e., $n=1$. This model is trained using metric learning.
  \item \texttt{Multi-1vs1}: The audio branch contains two CNN groups to obtain AEs of the user-liked/disliked track pair, respectively. Unlike \texttt{Basic-Binary}, this model uses a softmax function with categorical cross-entropy loss to maximize the positive relationships.
  \item \texttt{Multi-1vs4}: Similar to \texttt{Multi-1vs1}, it uses 4 user-disliked tracks as negative samples, i.e,  $n=4$.
  \item \texttt{Metric-1vs1}: This is our proposed model using metric learning with user-liked/disliked track pairs.
  \item \texttt{Metric-1vs4}: It is similar to \texttt{Metric-1vs1}, but uses 4 user-disliked tracks as negative samples.
\end{itemize}

\subsubsection{Results}
To efficiently compare the proposed model with other methods, we formed a sub-dataset including 12881 users and 52105 tracks. The split was performed on each user's liked and disliked tracks, respectively. Evaluation results with two choices of context duration are presented in Table \ref{subset-result}. We can observe all the models using the proposed UE outperforms \texttt{DCUE-1vs1}, which uses lookup-embeddings from user IDs as its UE. This shows our proposed UE can better capture users' music preference, and enable the audio branch to perform better parameter fitting to the audio content. A more representative AE is expected to be obtained. 

When context duration is set to 3 seconds, \texttt{Multi-1vs1} achieves the highest precision, while \texttt{Metric-1vs4} obtains the highest AUC score. Since a good music recommendation system should rank user-liked tracks ahead of the disliked ones, AUC is more suitable for evaluation. According to the AUC scores, our proposed models with metric learning achieve better performance than other classification-based models. We also increased the context duration to 10 seconds for training, leaving out \texttt{Multi} models for simplicity. Both precision and AUC scores are increased as shown in Table \ref{subset-result}. Considering the trade-off between model performance and computational efficiency, we decided \texttt{Metric-1vs1} with a context duration of 10 seconds as the best model for further use (denoted as \texttt{Metric-1vs1-10s}).

Finally, we evaluated \texttt{Metric-1vs1-10s} using the big dataset. Unlike the sub-dataset, we no longer share the same users across different splits for the big dataset, which includes 60 millions user-track interactions. The model can achieve a precision score of 0.773 and the AUC score of 0.849. With this trained model, AE can be obtained directly from any audio inputs.

\subsection{Evaluation on Genre Classification Tasks}
\label{sec:eval_genre}

\begin{table}[]
\centering
\begin{tabular}{@{}cccc@{}}
\toprule
SVM & \multicolumn{3}{c}{Classification Accuracy} \\ \midrule
 & GTZAN & FMA-small & MIREX08 \\
without AE & 0.7653 & 0.5918 & ----- \\
with AE & \textbf{0.8013} & \textbf{0.6148} & 0.7474 \\ \bottomrule
\end{tabular}
\caption{Evaluation results of SVM trained with or without AE in three datasets.}
\label{tab:genre-result}
\end{table}

Current genre classification systems typically use audio-based features as input. To efficiently obtain a genre classifier, we adopted transfer learning technique to put features extracted from the pre-trained models into a classifier based on an SVM. We used \textit{musicnn} \cite{musicnn} as the main pre-trained audio-based baseline. It is publicly available\footnote{\url{https://github.com/jordipons/musicnn}} and achieves state-of-the-art results on auto-tagging tasks. Variants of this model are dependent on the dataset it was trained on. We selected \texttt{MSD\_musicnn}, which was trained on roughly 200k tracks from the Million Song Dataset. Baseline features were then extracted from the \texttt{max\_pool} layer in this model. 

To explore the potential of AE obtained from the pre-trained \texttt{Metric-1vs1-10s}, we trained two genre classifiers. One used the baseline features only. The other one concatenated the baseline features and AE into a feature vector. In either case, features were decomposed via Principal Component Analysis (PCA) \cite{pca} into a vector of 128 dimensions. Finally, SVM was trained with radical kernels to classify the features by genres. Bandwidth for radial kernel was set to 1/128. The penalty parameter was set to 1.  

Experiments were conducted on two datasets: 
\begin{itemize}
  \item \textit{GTZAN}: 1000 tracks, 10 genres in \cite{gtzan}. We used a fault-filtered train/validation/test partitions\footnote{\url{https://github.com/jordipons/sklearn-audio-transfer-learning}} to obtain 443/197/290 tracks.
  \item \textit{FMA-small}:  8000 tracks, 8 balanced genres in \cite{fma_dataset}.  For each genre, we split it into 7:3 for training and testing.
\end{itemize}

We reported the averaged classification accuracy in Table \ref{tab:genre-result}. The SVM classifier trained with our AE presents better performance on both \textit{GTZAN} and \textit{FMA-small} datasets. We also participated in the MIREX 2020 genre classification tasks, and achieved an average accuracy of 0.7474 based on 3-fold artist-filtered cross-validation of the MIREX08 dataset\footnote{\url{https://www.music-ir.org/nema_out/mirex2020/results/act/mixed_report/summary.html}}. Plus, our proposed method shows comparable results in the K-pop related tasks. According to all the announced results of accuracy per class, we achieved the best place in classifying three genres: country, rap/hip-hop and K-pop ballad.

\section{Conclusion and Future Work}
In this paper, we presented a novel model to learn audio embeddings, which can be used as representations of new track releases to solve the cold-start problem in music recommendation systems. The model consists of a user branch and an audio branch. Through our study, we showed that using a more complete user listening data, including users' listening history and demographics, can yield a more informative user embedding from the user branch. This user embedding also helps to infer audio embeddings using metric learning in the audio branch. We evaluated the model in music recommendation and genre classification tasks. A better performance was reported in both tasks. This verified that our audio embedding is an effective representation of the audio input. 

For future works, a more detailed exploration of the model architecture for audio branch should be done. Also, the generalization ability of the learned audio embedding in different music information retrieval tasks has to be further investigated. We can make extensive use of offline metrics to iteratively improve our model. For the final determination of the effectiveness of the proposed audio embedding, online A/B testing on personalized recommendation of new track releases will be conducted.

\bibliographystyle{IEEEbib}
\bibliography{refs}

\end{document}